\documentclass[aps,prb,twocolumn,superscriptaddress,showpacs]{revtex4}

\usepackage{hyperref}
\usepackage[dvips]{graphicx}
\usepackage{amssymb,amsmath,amsfonts}
\usepackage[T1]{fontenc}
\newcommand{\alfa}{$\alpha_{\mathrm{hex}}$}

\newcommand{\doping}{$\Delta E_{F}$}

\begin{document}



\title{Fingerprints of Dirac points in first-principles calculations of scanning tunneling spectra of graphene on a metal substrate}


\author{J. S\l awi\'{n}ska}
\affiliation{Department of Theoretical Physics and Computer Science, University of Lodz, Pomorska 149/153, 90-236 Lodz, Poland}
\affiliation{Solid State Physics Department, University of Lodz, Pomorska 149/153, 90-236 Lodz, Poland}
\author{I. Zasada}
\affiliation{Solid State Physics Department, University of Lodz, Pomorska 149/153, 90-236 Lodz, Poland}


\vspace{.15in}

\begin{abstract}
Graphene physisorbed on a metal has its characteristic Dirac cones preserved in the band-structure, but the Fermi level of the system is shifted due to the interaction with the substrate. Based on density functional calculations with van der Waals corrections, we present a method to determine the position of the Dirac point with respect to the Fermi level from the measured scanning tunneling spectra (STS). It has been demonstrated that the dips in both simulated local density of states and in the observed $dI/dV$ profiles are indeed the fingerprints of the Dirac points. The type and the level of doping can be then inferred directly from the STS data without any additional experimental technique. Test calculations of graphene on a Cu(111) substrate have shown that the predicted position of the Dirac point is in close proximity to the experimental value reported in the recent studies. Moreover, simulations for graphene on a Pt(111) surface allow us to explain the apparent contradictions in the state-of-the-art experimental works.
\end{abstract}

\maketitle

\section{Introduction}
The unique properties of graphene \cite{review}, such as its ultrathin geometry \cite{novoselov_science} and high carrier mobility \cite{schwierz} make it a promising candidate for applications in future nanoelectronics, sensors and photonics \cite{nanoletters, polarizer}. Making complex devices requires the production of large enough high-quality graphene sheets to achieve scalability and benefit from exceptional electronic properties observed in the flat domains of the monolayer sheets \cite{dispersion_ru}. In practice, the large-area graphene has been grown epitaxially on transition metals \cite{iryd, ruten, platyna_sutter}, especially the chemical vapor deposition (CVD) method has been developed to synthesize graphene on copper \cite{cu_science} and gold \cite{gold_cvd} substrates, although their catalyst's role is still not well understood \cite{cu_carbon}. Due to the high quality and transferability of the samples prepared on metallic surfaces\cite{nature_revised}, much theoretical \cite{holendrzy_prl, holendrzy, interface_xu, vanin, comparative, nasza6, platyna_hiszpanie} and experimental effort \cite{cu_sts, cu_revised1, cu_revised2, platyna_sutter, platyna_sts, klusek} has been devoted to shed light on the mechanisms of graphene-metal interactions as well as on the modifications of its electronic properties. In particular, the doping effect \cite{nasza6} in physisorbed graphene has been widely studied both theoretically and experimentally by using a number of techniques, as for example density functional theory (DFT), scanning tunneling microscopy (STM) and angle-resolved photoemission spectroscopy (ARPES).

Scanning probe methods \cite{stm_hofer2} have developed into a powerful tool for the determination of the structure of surfaces and interfaces, thus it is especially suitable for studies of graphene deposited on conductive substrates. The STS mode allows us to record the first derivative of the tunneling current with respect to the bias voltage which is the measure of the surface density of states at every arbitrarily fixed point. The electronic structure is probed locally \cite{sts_theory2}, thus the presence of heterogeneity at different places on the surface or periodically repeated domains can be identified. Moreover, the information as to how the metal surface states are modified by the adsorption of graphene sheet can be obtained after careful analysis.

According to early local density approximation (LDA)- based DFT studies \cite{holendrzy} of a graphene/Cu(111) system, the physisorption leads to the change in the location of Dirac point, since the electrons are donated by a substrate to graphene (n-type doping of -0.17 eV). In a recent experimental study of graphene epitaxially grown on a Cu(111) surface the results of STS measurements are reported \cite{cu_sts}. The dip in the presented STS profile (Fig.3C in Ref.\onlinecite{cu_sts}) at approximately -0.35 eV is associated with the Dirac point shifted below the Fermi level, which qualitatively confirms previous theoretical results. However, the interpretation of differential tunneling spectra is not straightforward due to the influence of the tip states \cite{sts_theory2}. Thus, as suggested in Ref.\onlinecite{cu_sts} only a combination of STS with a secondary experimental technique could give a direct evidence that a dip reflects the local density of states at the Dirac point of graphene. ARPES measurement could confirm the energy position of the Dirac point, however sufficiently large monocrystalline samples are required to find specific paths inside the Brillouin zone of the system. On the other hand, the new methods of Dirac point mapping have been proposed, such as following the gate-dependent position of adjacent dips in the tunable system \cite{giant_phonon} or analysis of Fourier transforms of STM maps revealing a linear dispersion relation for states above and below the Fermi energy \cite{nature_origin}. They are, however, either not applicable for conductive substrate or at least hardly achievable without special equipement such as low temperature (LT) STM with lock-in detection.  

In contrast, the theoretical predictions and the experimental findings for the graphene/Pt(111) system seem to be in an excellent agreement. The adsorption distance of about 3.30 \AA\ is both predicted by DFT calculations\cite{holendrzy} and measured by low-energy electron diffraction (LEED) technique.\cite{platyna_sutter} The doping level is estimated to approximately 0.33 eV by DFT calculations\cite{holendrzy} while ARPES measurements gave the value of 0.3 eV.\cite{platyna_sutter} However, ARPES can determine only occupied states: in the case of the Fermi level shift downward with respect to the Dirac point, the value of doping must be extrapolated. In this view, the main advantage of STS is its ability to record information about unoccupied part of the spectrum, which allows us to measure doping directly. Recent experimental studies revealed two types of spectra for the graphene/Pt(111) system: the one with clearly visible onset of a Pt(111) surface state\cite{platyna_sts} and the second one with global minimum corresponding to the Dirac point\cite{platyna_hiszpanie} (in agreement with calculated PDOS). We propose the possible explanation of this difference due to analysis of metallic surface states in terms of the LDOS above the surface.

All these indicate that linking theoretical predictions and experimental findings needs an approach independent of the limitations of measurement and preparation methods. In this paper, we report on the first-principles simulations of differential spectra that can be directly compared with measured $dI/dV$ profiles of graphene on a metal surface. As the DFT method provides the information about the band structure, the dip in calculated LDOS above the surface can be unambiguously linked to the energy position of Dirac points in the dispersion relation. The examples of graphene on a Cu(111) and Pt(111) substrates have been carefully elaborated to test the presented method.

The paper is organized as follows: details of DFT calculations are described in Sec. II, whereas the idea of a proposed method as well as two examples of its application are presented in Sec. III. Final remarks are given in Sec. IV.

\section{DFT calculations}
All DFT calculations have been performed using \textsc{vasp} package \cite{vasp1, vasp2} equipped with the projector augmented wave (PAW) method \cite{paw1, paw2} for electron-ion interactions. The exchange-correlation energy is calculated using the generalized gradient approximation within the Perdew, Burke and Ernzerhof (PBE) \cite{pbe} parametrization scheme. Long-range dispersion corrections have been taken into account within a DFT-D2 approach of Grimme \cite{grimme}, as implemented in the latest version of \textsc{vasp} \cite{bucko}. The pair interactions up to a radius of 12 \AA\, have been included in the calculations and the global scaling factor $s_{6}$ has been set to 0.75 due to the choice of the PBE functional. The dispersion coefficients $C_{6}$ and van der Waals radii $R_{0}$ for C and Cu atoms are defined in the code, according to the suggestion of Grimme \cite{grimme}, while the $C_{6}$ and $R_{0}$ coefficients of Pt have been set to 20 Jnm$^{6}$/mol and 1.9\AA, respectively, as in Ref. \onlinecite{platyna_hiszpanie}.\cite{private_correspondence} The electronic wave functions have been expanded in a plane-wave basis set of 400 eV, while the electronic self-consistency criterion has been set to $10^{-7}$ eV. 

\begin{figure}
\includegraphics[width=0.45\textwidth]{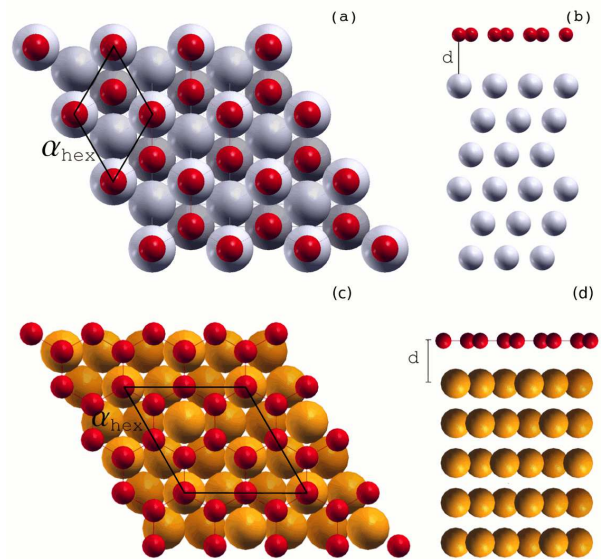}
\caption{\label{geometria} (Color online) Top view (a) and side view (b) of adsorption geometry of graphene on a Cu(111) substrate. Carbon atoms are denoted as red (darker) balls, copper atoms as gray balls. Top view (c) and side view (d) of adsorption geometry of graphene on a Pt(111) surface. Platinum atoms are denoted as yellow (lighter) balls. Parallelograms define the unit cells.}
\end{figure}
The (111) surfaces of metals are simulated by a periodic slab geometry. Supercells containing six (five) layers of Cu (Pt) atoms and a graphene sheet adsorbed on top are separated by the vacuum thickness of at least 20 (15) \AA. The typical in-plane adsorption geometry is shown in Fig.\ref{geometria}(a),(b) and Fig.\ref{geometria}(c),(d) for Cu(111) and Pt(111) substrate, respectively. The graphene-metal interface is modeled by direct matching of a graphene's and a substrate's unit cells, although in the real systems the lattice-mismatch leads to the Moir\'{e} pattern on the surface. This artificially generated strain tends to affect the surface relaxation and the electronic properties, which were discussed in Ref. \onlinecite{nasza6}. Thus, for the graphene/Cu(111) system we have performed calculations for both configurations: with the in-plane lattice parameter \alfa\, equal to the PBE+D2-optimized value for graphene 2.466 \AA\, as well as with the lattice constant adapted to that of a metal (two cases were considered: PBE+D2 optimized \alfa\, =2.524 \AA\, and an experimental value of 2.56 \AA). There exist also three inequivalent stacking orders of graphene and Cu layers. The carbon atoms can lie above metal atoms in layers 1 and 3 (known as the top-fcc configuration, see Fig.\ref{geometria}), 1 and 2 (a top-hcp) or 2 and 3 (a hcp-fcc). The previous LDA study \cite{interface_xu} identified the top-fcc configuration as the most stable one, which was also confirmed by the present vdW calculations (total energies of top-hcp and hcp-fcc configurations are higher than top-fcc one by 2.3 meV and 13.1 meV, respectively). In view of the above test calculations, in the case of graphene/Pt(111) system we have used only one value of in-plane lattice paramater \alfa\, as well as $2\times2$ graphene's unit cell in configuration defined in Fig. \ref{geometria}(c). During all structure relaxations the metal in top two layers as well as all carbon atoms have been allowed to move. Total energies were converged to within $10^{-6}$ with respect to the ionic steps.

The problem of accurate Brillouin zone sampling is of particular importance for simulations of local density of states above the surface. The simulations of STM images are not computationally demanding, but the resolution of the calculated STS profiles is determined by the accuracy of the band-structure maps. For example, the identification of the onset of a noble metal surface requires at least one k-point every few meV for a specific band which means a few thousands of k-points in the Brillouin zone \cite{hofer_differential}. During the relaxation and the self-consistent run we applied the tetrahedron scheme \cite{tetrahedron} and the $\Gamma$-centered 36$\times$36$\times$1 and 24$\times$24$\times$1 k-point meshes for small and large supercell, respectively, whereas in the final local density of states calculations, a dense $336\times336\times1$ ($192\times192\times1$ for larger supercell) $\Gamma$-centered k-points mesh and the Gaussian smearing of width $\sigma$ =0.05 eV have been used.

\section{Identification of Dirac points in STS data}
The Tersoff-Hamann model of tunneling, due to its simplicity and qualitative reliability, is implemented in \textsc{vasp} as well as nearly every other DFT code to make simulations of pseudo-STM topography images. In this approach the tunneling current is proportional to the local density of states at the position of the STM tip \cite{stm_hofer2}:
\begin{displaymath}I(\mathbf{R}) \sim \sum_{E_{n}>E_{F}-eV_{\mathrm{bias}}}^{E_{n}<E_{F}}|\psi(\mathbf{R}, E_{n})|^{2}\end{displaymath}

Provided that the k-points mesh is sufficiently fine, the reliable pseudo-STS profiles can be obtained by an evaluation of the constant-height charge images for several values of bias voltage followed by their numerical differentiation. The local density of states is automatically calculated in every single point above the surface and the choice of a spectrum in the particular position could correspond to the STS data taken under open loop conditions at the fixed point.

\subsection{Cu(111) substrate: n-type doping of graphene}
\begin{figure}
\includegraphics[scale=1.0]{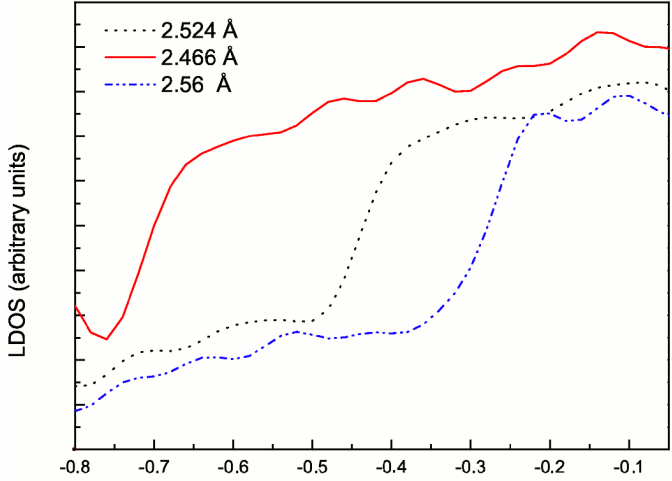}
\caption{\label{sieci} (Color online) The simulated LDOS spectra of a clean Cu(111) surface for different choices of the in-plane lattice parameter \alfa.}
\end{figure}

The local densities of states of a clean perfect Cu(111) surface are shown in Fig. \ref{sieci}. We have evaluated the spectra for all three considered choices of the in-plane lattice paramater \alfa. Their overall shape is similar to those presented in the previous theoretical studies \cite{hofer_differential, theory2} as well as to the data obtained in a number of experiments \cite{experiment1, experiment2, experiment3}. The step-shaped onsets of the surface state can be easily observed at bias voltages $U_{1}=-700$ mV, $U_{2}=-440$ mV and $U_{3}=-280$ mV for the values of lattice constants $\alpha_{\mathrm{hex}}^{1}$=2.466, $\alpha_{\mathrm{hex}}^{2}$=2.524 and $\alpha_{\mathrm{hex}}^{3}$=2.56 \AA, respectively. It should be noted that the oscillations present in the profiles originate from a discrete k-points mesh. According to the discussion presented in Ref. \cite{hofer_differential}, the signature of a surface electronic structure should be very sensitive to the choice of the lattice constant. Indeed, the experimental position of the surface state is reproduced only for a value of the lattice constant $\alpha_{\mathrm{hex}}^{2}$ optimized within the PBE+D2 approach. This appears to be in contrast with previous theoretical results based on a Perdew-Wang (PW91) parametrization, where setting the \alfa\, to an experimental value was needed to obtain the quantitatively correct data. The discrepancies seem to be caused by using of a different parametrization approach and will be widely discussed in a separate study. The problem of the surface onset shift is especially important for modeling the graphene-metal interface where adjusting the substrate's lattice constant to graphene is usually required \cite{nasza6}. It means that a correct theoretical determination of subtle features such as experimentally observed suppression of a Cu(111) surface state in the presence of graphene \cite{cu_sts} could be hardly achievable within a $1\times1$ supercell approach.

The bonding of graphene to a copper surface is rather weak and the charactersitic Dirac cones are nearly preserved in the band-structure. Both LDA \cite{hbn} and the present DFT-D2 calculations predict a tiny band-gap of the order of 10 meV in the case of the matched unit cells, but in a real incommensurate interface it should disappear. It is worthwile to notice the interesting fact that the presence or lack of this gap is closely related to the symmetries (hexagonal or trigonal) being preserved in the system. Dirac fermions can exist even in chemisorbed graphene provided that the defects are distributed symmetrically. This problem is widely and elegantly discussed in Ref. \onlinecite{prbrevised_gorsza} in terms of symmetries, group theory and tight binding Hamiltonians. It has been also shown by means of DFT calculations\cite{prbrevised_lepsza} that despite functionalization through dopants or vacancy defects, the presence of Dirac fermions may be recovered for superstructures and nanomeshes with specific symmetry. In the case of Cu(111) substrate, however, real weak interaction with graphene has been confirmed experimentally.

\begin{figure}
\includegraphics[width=\columnwidth]{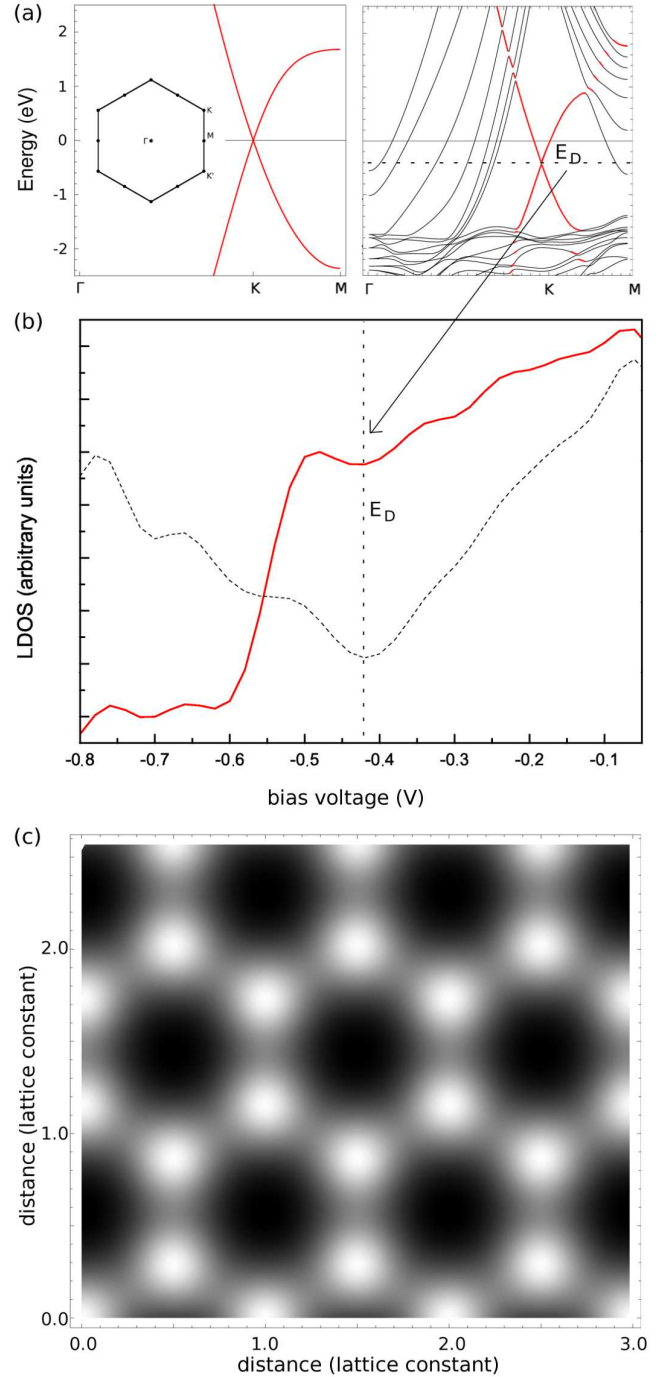}
\caption{\label{domieszkowanie} (Color online) (a) Left-hand panel: The band-structure of freestanding graphene. The inset shows a Brillouin zone with the $\Gamma$, K, and M high-symmetry points. Right-hand panel: The electronic structure of graphene on Cu(111). (b) The calculated LDOS profile above the surface of a graphene/Cu(111) system for a sample-tip distance larger than 2 \AA\, (solid line) and equal to about 1 \AA\,(dashed line). The parameter \alfa\, has been set to 2.466 \AA. The zero energy is at a Fermi level. (c) Simulated constant-height STM image corresponding to calculated LDOS presented above. Bias voltage is set to U =-100 mV. Light spots are in the positions of carbon atoms.}
\end{figure}

The physisorption of graphene on a Cu(111) substrate causes a Fermi level shift upward \cite{holendrzy}. In Fig.\ref{domieszkowanie} (a) we present the energy dispersion relation of graphene/Cu(111) system for a value of \alfa\,=2.466 \AA. The similar band-structures have been obtained for other values of the lattice constant, i.e. 2.524 \AA\, and 2.56 \AA. The optimized adsorption distances are about 3.06 \AA\, 2.98 \AA\, and 2.88 \AA, respectively. It is worthwhile to note that DFT-D2 tends to overestimate binding energies, e.g. for the considered system binding energy per carbon atom equals 117 meV, which is very close to the values for physisorbed graphene reported in a recent DFT-D2 study\cite{nasza6}, but significantly differs from 38 meV calculated within vdw-DF approach by Vanin \textit{et al.}\cite{vanin}. Next, we define the doping level as \doping= $E_{D}-E_{F}$, i.e. a difference between Dirac point energy and a Fermi level. The values of doping levels are equal to -0.427 eV, -0.584 eV and -0.686 eV for the considered lattice parameters, respectively. The problem of changing in doping with the variation of the in-plane lattice parameter has been discussed in details in Ref. \onlinecite{nasza6}: it seems that the most reliable value of doping is provided by the choice of in-plane lattice parameter matched to graphene's one. The differences in doping can be also described in terms of the charge transfer, in particular evaluated using the Bader scheme\cite{bader1, bader2, bader3}. The charge transfer $\delta Q$ equals to -7.4, -24.1 and -31.3 $\times 10^{-3}e$, respectively, where negative sign indicates electron transfer to graphene. The increase in $\delta Q$ with larger values of \alfa\, seems to be related to decrease in adsorption distance. Moreover, the adsorption distances are usually underestimated\cite{aromatic} by DFT-D2, thus the calculated doping level can be slightly below the measured values.

The STM experiment of graphene on a Cu(111) substrate revealed a Moir\'{e} pattern present on the surface \cite{cu_sts}. It means that the length of graphene bonds is preserved, which seems to be common for all weakly binding systems. Moreover, it has been previously demonstrated for a similar system, that the most reliable values of doping are obtained when a unit cell is adjusted to graphene's lattice constant. For example, in case of graphene - gold interface, stretching of bonds leads to the value of doping which is in disagreement with the experimental data \cite{klusek}. The choice of \alfa = 2.466 \AA\, should then lead to the most accurate description of the graphene's electronic properties.

In Fig. \ref{domieszkowanie} (b) we present a LDOS spectrum for a graphene/Cu(111) interface simulated with the \alfa =2.466 \AA\, (solid line). Since the position of the Dirac point is known from the band-structure (see Fig. \ref{domieszkowanie} (a)), we can directly map it onto an STS profile. It means that a dip at -0.427 eV in the spectrum (Fig. \ref{domieszkowanie} (b), solid line) can be associated with the Dirac point in the energy dispersion, which provides an unambiguous link between experimental STS profiles and the calculated band-structures. In this light, the local minimum observed at -0.35 eV in the spectrum (Fig.3C in Ref. \cite{cu_sts}) should be indeed interpreted as a signature of a Dirac point. The measured value of doping is quite close to the one predicted in this study. The difference of about 80 meV might be explained by the effect of phonon-mediated inelastic tunneling as recently discussed in Ref. \cite{giant_phonon}. This phenomenon could cause a systematic shift of a Dirac point dip toward the Fermi level by about few tens of meV. Another reason for a doping level shift may be connected with the adsorption distance underestimated in the DFT-D2 approach.

Despite the oscillations related to the finite resolution of the spectrum presented in Fig. \ref{domieszkowanie} (b), a Dirac point dip can be still easily recognized. However. if the profile is evaluated using the minimal required k-points mesh the Dirac point can be hardly identified. One may additionally confirm its location by decreasing the tip-sample distance, i.e. approaching to the surface (Fig. \ref{domieszkowanie} (b), dashed line). At the distances of the order of 1 \AA\, above the surface, the graphene states dominate in the spectrum and the minimum can be unambiguously recognized.

It should be also noted that the onset of a Cu(111) surface state after the deposition of graphene (Fig.\ref{domieszkowanie} (b), solid line) is shifted toward the Fermi level with respect to the position predicted in the calculations for a clean surface done using the same lattice-constant parameter \alfa=2.466 \AA\, (Fig. \ref{sieci} , solid line). It appears to be connected with the changes in the work function upon adsorption of the carbon atoms.  

\subsection{Pt(111) substrate: p-type doping of graphene}
\begin{figure}
\includegraphics[width=\columnwidth]{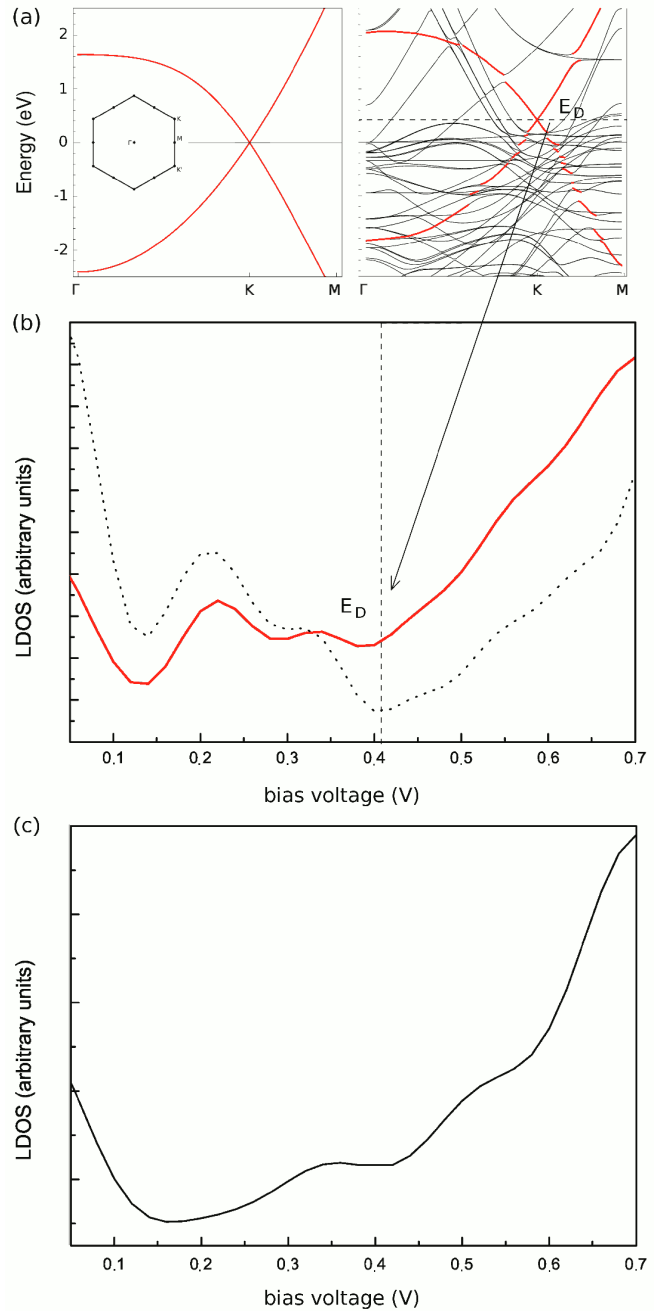}
\caption{\label{platyna} (Color online) (a) Left-hand panel: The band-structure of freestanding graphene in 2$\times$2 unit cell. The inset shows a Brillouin zone with the $\Gamma$, K, and M high-symmetry points. Right-hand panel: The electronic structure of graphene on Pt(111). (b) The calculated LDOS profile above the surface of a graphene/Pt(111) system for a sample-tip distance larger than 5 \AA\, (solid line) and equal to about 3 \AA\,(dashed line). (c) The simulated LDOS profile of a clean Pt(111) substrate. It agrees well with previous theoretical and experimental spectra\cite{pt_pristine}. The zero energy is at a Fermi level.}
\end{figure}
In the case of the Pt(111) substrate, we have restricted our analysis to only one fixed lattice constant (see Ref.\onlinecite{nasza6} for wider discussion). In Fig. 4(a) we present the band-structure of graphene interacting with Pt(111): we can observe that, in accordance with previous results\cite{holendrzy, platyna_hiszpanie}, Dirac cones are preserved and shifted upward with respect to the Fermi level, which results in p-type doping of graphene (0.41 eV). In spite of the fact that we used only $2\times2$ unit cell of graphene the value of doping agrees with that presented in Ref.\onlinecite{platyna_hiszpanie}, where the Moir\'{e} pattern was taken into account. It should be noted that the band-structure measured by ARPES(\onlinecite{platyna_sutter}) also confirms the presence of Dirac fermions in this lattice-mismatched system.

In Fig. 4(b) we present a LDOS profiles of the graphene/Pt(111) system and, for comparison, spectrum of the pristine Pt(111) surface  (Fig. 4 (c)). Similarly as in the case of graphene/Cu(111) system, we can associate the position of the Dirac point in the band-structure $E_{D}$ with a dip at energy 0.41 eV in calculated LDOS spectra above the surface. Linking two types of data allows to recognize the fingerprints of Dirac points in simulated and measured $dI/dV$ spectra. However, one can notice at least two differences comparing with the graphene/Cu(111) system: (i) the surface states of Pt(111) appear to be better pronounced: even in the presence of global minimum responsible for Dirac point (dashed line in Fig. 4(b)) the surface state of platinum can be easily recognized; (ii) it is not unambiguous as to which of two dips in the LDOS curve (solid line in Fig. 4(b)) reflects the Dirac point: one of them may be induced by Pt surface state; in this context the evaluation of LDOS close to the surface is needed (dashed line).

In this light, it is not surprising that contradictory experimental results have been recently reported. In STS data presented in Ref.\onlinecite{platyna_sts} (Fig.1B therein) there is no signature that could be associated with the presence of Dirac point in any particular position. It is supposed\cite{platyna_sts} that this is due to Pt(111) surface state features which are still seen in the spectra. We suggest that in such cases only direct and exact comparison with simulated LDOS might allow one to identify the Dirac point. On the other hand, $dI/dV$ spectrum shown in Ref.\onlinecite{platyna_hiszpanie} (Fig.1(e) therein) corresponds rather to the dashed line in Fig 4 (b) in this present paper as well as to PDOS profile calculated in Ref.\onlinecite{platyna_hiszpanie}. In this case the surface states are hardly visible which allows one to associate the global minimum to a Dirac point. The absence or presence of surface state may be related to preparation technique or details of the STS experiment.

\section{Final remarks}

In summary, we have performed DFT-D2 calculations for graphene interacting with Cu(111) and Pt(111) substrates and evaluated the corresponding LDOS spectra above the surface. The fingerprints of Dirac points present in the band structure are easily recognized in the simulated STS profiles, which enables a straightforward comparison of theoretical and experimental data. The approach can be successfully used even for the systems where the Dirac point and the surface state onset seem to be overlapped. The doping of graphene on Cu(111) predicted in this study is in a better agreement with the measured value than the previous LDA results. It seems that using the dispersion corrections in DFT calculations is essential for accurate determination of electronic characteristics and demonstrates that vdW forces are responsible for properties of the graphene-metal interfaces. Moreover, the simulations for the graphene/Pt(111) system explain the difficulties with identification of Dirac points in the experimental STS spectra.

\begin{acknowledgments}
We thank Z. Klusek and P. Dabrowski for helpful discussions. This work is financially supported by Polish Ministry of Science and Higher Education in the frame of Grant No. N~N202~204737. One of us (J.S.) acknowledges support from the European Social Fund implemented under the Human Capital Operational Programme (POKL), Project: D-RIM. Part of the numerical calculations reported in this work have been performed at the Interdisciplinary Center for Mathematical and Computational Modelling (ICM) of the University of Warsaw within the Grant No. G44\,-\,2 as well as at the Computer Center of Technical University of Lodz using project PLATON infrastructure. Figure 1 was prepared using the \textsc{xcrysden} program. \cite{kokalj}
\end{acknowledgments}

\end{document}